\documentclass[useAMS,usenatbib]{mn2e}
\usepackage{epsfig}
\usepackage{color}
\usepackage{subfigure}

\newcommand{\goodgap}{\hspace{\subfigtopskip} \hspace{\subfigbottomskip}}

\title[Systematics in the GRBs Hubble diagram]{Systematics in the Gamma Ray Bursts Hubble diagram}
\author[V.F. Cardone et al.]{V.F. Cardone$^{1}$\footnote{Corresponding author\,: {\tt winnyenodrac@gmail.com}}, M. Perillo$^{2,3}$, S. Capozziello$^{4}$\\
$^1$I.N.A.F. - Osservatorio Astronomico di Roma, via Frascati 33, 00040\,-\,Monte Porzio Catone (Roma), Italy \\
$^2$Dipartimento di Fisica ``E.R. Caianiello", Universit\`{a} di Salerno, Via Ponte Don Melilo, 84081 Fisciano (Sa), Italy \\
$^3$ I.N.F.N. - Sez. di Napoli, Compl. Univ. Monte S. Angelo, Ed. G, Via Cinthia, 80126 Napoli, Italy \\
$^4$Dipartimento di Scienze Fisiche, Universit\`{a} degli Studi di Napoli "Federico II",
Complesso Universitario \\ di Monte Sant'Angelo, Edificio N, via Cinthia, 80126 - Napoli, Italy}

\date{Accepted xxx, Received yyy, in original form zzz}

\begin{document}

\maketitle

\begin{abstract}

Thanks to their enormous energy release which allows to detect them up to very high redshift, Gamma Rays Bursts (GRBs) have recently attracted a lot of interest to probe the Hubble diagram (HD) deep into the matter dominated era and hence complement Type Ia Supernoave (SNeIa). However, lacking a local GRBs sample, calibrating the scaling relations proposed as an equivalent to the Phillips law to standardize GRBs is not an easy task because of the need to estimate the GRBs luminosity distance in a model independent way. We consider here three different calibration methods based on the use of a fiducial $\Lambda$CDM model, on cosmographic parameters and on the local regression on SNeIa. We find that the calibration coefficients and the intrinsic scatter do not significantly depend on the adopted calibration procedure. We then investigate the evolution of these parameters with the redshift finding no statistically motivated improvement in the likelihood so that the no evolution assumption is actually a well founded working hypothesis. Under this assumption, we then consider possible systematics effects on the HDs introduced by the calibration method, the averaging procedure and the homogeneity of the sample arguing against any significant bias. We nevertheless stress that a larger GRBs sample with smaller uncertainties is needed to definitely conclude that the different systematics considered here have indeed a negligible impact on the HDs thus strengthening the use of GRBs as cosmological tools.

\end{abstract}


\section{Introduction}

The observational evidences accumulated in these less than fifteen years, from the anisotropy and polarization spectra of the cosmic microwave background radiation (CMBR, \cite{Boom00,QUAD09,WMAP7}, the large scale structure traced by galaxy redshift surveys \citep{D02,P02,Sz03,H03}, the matter power spectrum \citep{Teg06,P07} with the imprints of the Baryonic Acoustic Oscillations (BAO, \cite{Eis05,P10}) and the Hubble diagram of TypeIa Supernovae (SNeIa, \cite{Union,H09,SNeIaSDSS}), definitely support the cosmological picture of a spatially flat universe with a subcritical matter content $(\Omega_M \sim 0.3)$ and undergoing a phase of accelerated expansion. While the observational scenario is now firmly established, the search for the motivating theory is, on the contrary, still in its infancy notwithstanding the many efforts and the plethora of models proposed along these years. Ironically, the problem here is not the lack of a well established theory, but the presence of too many viable candidates, ranging from the classical cosmological constant \citep{CPT92,SS00} to scalar fields \citep{PR03,Copeland06} and higher order gravity theories \citep{CF08,FS08,ND08,dFT10}, all of them being more or less able to fit the available data.

As often in science, adding further data is the best strategy to put order in the confusing abundance of theoretical models. In particular, pushing the observed Hubble diagram to higher redshift would allow to trace the universe background evolution up to the transition regime from the dark energy driven speed up to the decelerated matter dominated phase. Moreover, being the distance modulus $\mu(z)$ log\,-\,linearly related to the luminosity distance and depending this latter on the dark energy equation of state through a double integration, one has to go to large $z$ in order to discriminate among the prediction of different models when these predict similar $\mu(z)$ curves at lower redshift. Unfortunately, SNeIa are not ideally suited to this task with their present day Hubble diagram going back to $z_{max} \sim 1.4$ and not extending further than $z \simeq 2$ even for excellent space based projects such as SNAP \citep{SNAP}.

Thanks to their enormous almost instantaneous energy release, Gamma Ray Bursts (GRBs) stand out as ideal candidates to go deeper in redshift, the farthest one yet being at $z = 8.3$ \citep{Salvaterra2009}. The wide range spanned by their peak energy makes them everything but standard candles, but the existence of many observationally motivated correlations between redshift dependent quantities and rest frame properties \citep{Amati08,FRR00,N00,G04,liza05} offers the intriguing possibility of turning GRBs into standardizeable candles just as SNeIa. The use of these scaling relations allows to infer the GRB distance modulus with an error mainly depending on the correlation intrinsic scatter.  Combining the estimates from different correlations, Schaefer (2007) first derived the GRBs Hubble diagram (hereafter, HD) for 69 objects, while Cardone et al. (2009) used a different calibration method and add a further correlation to update the GRBs HD. Many attempts on using GRBs as cosmological tools have since then been performed (see, e.g., Firmani et al. 2006, Liang et al. 2009, Qi \& Lu 2009, Izzo et al. 2009 and refs. therein) showing the potential of GRBs as cosmological probes.

It is worth stressing that the possibility offered by GRBs to track the HD deep into the matter dominated epoch does not come for free. Two main problems are actually still to be fully addressed. First, lacking a local GRBs sample, all the above correlations have to be calibrated assuming a fiducial cosmological model to estimate the redshift dependent quantities. As a consequence, the so called circularity problem comes out, that is to say one wants to use GRBs scaling relations to constrain the underlying cosmology, but needs the underlying cosmology to get the scaling relations. Different strategies have been proposed to break this circularity so that our first aim here is to investigate whether they are indeed viable solutions and to what extent the residual problem impact the HD derivation.

A well behaved distance indicator should be not only visible to high $z$ and possess scaling relations with as small as possible intrinsic scatter, but its physics should be well understood from a theoretical point of view. On the contrary, there is up to now not any definitive understanding of the GRBs scaling relations so that, as a dangerous drawback, one can not anticipate whether the calibration parameters are redshift dependent or not. Lacking any theoretical model, we will therefore address here this problem in a phenomenological way adopting two different parameterizations for their evolution with $z$ allowing for a change in the zeropoint only or in both zeropoint and slope. It is worth noting that such an analysis has been severely hampered before by the low statistics in the GRBs sample, a problem that we partially overcome here thanks to the use of the 115 GRBs catalog recently assembled by Xiao \& Schaefer (2010).

The plan of the paper is as follows. In Sect.\,2, we briefly review the different GRBs 2D correlations known insofar and present the Bayesian motivated method we will use in the following to calibrate them using three different approaches to estimate the GRBs luminosity distances. The constraints thus obtained and their dependence on the adopted luminosity distance determination are discussed in Sect.\,3, while, in Sect.\,4, we address the problem of their dependence on the redshift. The issues related to the HD derivation are then discussed in Sect.\,5 where we consider the impact on the HD of the luminosity distance estimate method, of the the averaging procedure and the homogeneity of the sample. A summary of the main results and a discussion of them is finally given in Sect.\,6, while some complementary material is presented in Appendix.

\section{GRBs scaling relations}

It is instructive to start this analysis considering the general case of two observable quantities $(x, y)$ related by a power\,-\,law relation which, in a log\,-\,log plane, reads

\begin{equation}
\log{y} = a \log{x} + b  \ .
\label{eq: loglog}
\end{equation}
Calibrating such a relation means determining the slope $a$, the zeropoint $b$ and the scatter $\sigma_{int}$ of the points around the best fit relation. In order to be useful for distance determination, one of the two quantities, say $x$, should refer to a directly observed quantity, while the other one must depend on the redshift $z$. Setting $y = \kappa d_L^2(z)$ with $\kappa$ a directly measurable redshift independent quantity and $d_L(z)$ the luminosity distance, we get

\begin{displaymath}
\log{y} = \log{\kappa} + 2 \log{d_L(z)} = a \log{x} + b
\end{displaymath}
so that one can then estimate the distance modulus as\,:

\begin{displaymath}
\mu =  25 + 5 \log{d_L(z)} = 25 + (5/2) \left ( a \log{x} + b - \log{\kappa} \right ) \ .
\end{displaymath}
In order to perform such an estimate, a two steps procedure has to be implemented. First, one has to select a sample of low redshift $(z \le 0.01)$ objects with known distance and fit the scaling relation to the $(x, y)$ data thus estimated to infer the calibration parameters $(a, b, \sigma_{int})$. Second, one has to assume that such calibration parameters do not change with the redshift so that a measurement of $(x, \kappa, z)$ and the use of the above scaling relation, with $(a, b, \sigma_{int})$ the same at all $z$, are sufficient to infer the distance modulus.

Both these steps are daunting tasks for GRBs because of the lack of low redshift objects to be used in the calibration procedure. In order to overcome this problem, different strategies can be implemented, but their impact on the final derivation of the GRBs Hubble diagram have not been investigated in detail which is our aim here.

\subsection{2D empirical correlations}

Their high luminosity offering the possibility to be detected at very large $z$ makes GRBs promising candidates to trace the Hubble diagram in the matter dominated era so that a great effort has been devoted during the late years to look for reliable and narrow empirical correlations. We limit here our attention only to two dimensional (hereafter, 2D) correlations since they can be investigated relying on a larger number of GRBs. These involve a wide range of GRBs properties related to both the energy spectrum and the light curve which are then correlated with the isotropic luminosity $L$ or the emitted collimation corrected energy $E_{\gamma}$. Neither $L$ nor $E_{\gamma}$ are directly measurable quantities since they depend on the luminosity distance $d_L(z)$. Indeed, it is\,:

\begin{equation}
L = 4 \pi d_L^2(z) P_{bolo} \ ,
\label{eq: lpbolo}
\end{equation}

\begin{equation}
E_{\gamma} = 4 \pi d_L^2(z) S_{bolo} F_{beam} (1 + z)^{-1}  \ ,
\label{eq: egamma}
\end{equation}
where $P_{bolo}$ and $S_{bolo}$ are the bolometric peak flux and fluence, respectively, while $F_{beam} = 1 - \cos{(\theta_{jet})}$ is the beaming factor with $\theta_{jet}$ the rest frame time of achromatic break in the afterglow light curve. Note that the bolometric quantities are related to the observed ones as \citep{S07}\,:

\begin{equation}
P_{bolo} = P  \ {\times} \ \frac{\int_{1/(1 + z)}^{10^4/(1 + z)}{E \Phi(E) dE}} {\int_{E_{min}}^{E_{max}}{E \Phi(E) dE}} \ ,
\label{eq: defpbolo}
\end{equation}

\begin{equation}
S_{bolo} = S \ {\times} \ \frac{\int_{1/(1 + z)}^{10^4/(1 + z)}{E \Phi(E) dE}} {\int_{E_{min}}^{E_{max}}{E \Phi(E) dE}} \ ,
\label{eq: defsbolo}
\end{equation}
with $P$ and $S$ the observed peak energy and fluence and $(E_{min}, E_{max})$ the detection thresholds of the observing instrument. We model the energy spectrum $\Phi(E)$ as a smoothly broken power\,-\,law, i.e. \citep{Band93}\,:

\begin{equation}
\Phi(E) = \left \{
\begin{array}{ll}
\displaystyle{A E^{\alpha} \exp{\left [ - \frac{(2 + \alpha) E}{E_{peak}} \right ]}}
& \displaystyle{\frac{E}{E_{peak}} \le \frac{\alpha - \beta}{2 + \alpha}} \\
~ & ~ \\
\displaystyle{B E^{\beta}} & {\rm otherwise}
\end{array}
\right .
\label{eq: band}
\end{equation}
The GRBs 2D correlations are obtained setting $y = L$ or $y = E_{\gamma}$ in Eq.(\ref{eq: loglog}), while different choices are available for $x$. In particular, we will consider the following possibilities\,: (i.) $x = E_{peak}(1 + z)/300$ with $E_{peak}$ the peak energy (in keV) of the fluence spectrum; (ii.) $x = \tau_{lag}(1 + z)^{-1}/0.1$ with $\tau_{lag}$ (in s) the time offset between the arrival of the low and high energy photons; (iii.) $x = \tau_{RT} (1 + z)^{-1}/0.1$ with $\tau_{RT}$ (in s) the shortest time over which the light curve rises by half the peak flux of the burst; (iv.) $x = V (1 +z)/0.02$ with $V$ the variability which quantifies the smoothness of the light curve itself. Note that all these quantities are expressed in the GRB rest frame which motivates the redshift term, while the further scaling constant is introduced to minimize the correlation among errors.

The combination of $x$ and $y$ gives rise to the different correlation we will consider\footnote{Hereafter, we will refer to the correlation obtained setting $x = X$ and $y = Y$ in Eq.(\ref{eq: loglog}) as the {\it $Y$\,-\,$X$ correlation} or {\it scaling law}.}, namely the $E_{\gamma}$\,-\,$E_{peak}$ \citep{G04,G06}, the $L$\,-\,$E_{peak}$ \citep{S03,Yo04}, $L$\,-\,$\tau_{lag}$ \citep{N00}, $L$\,-\,$\tau_{RT}$ \citep{S07} and $L$\,-\,$V$ \citep{FRR00,R01,S07}. These five correlations has been used in Schaefer (2007, hereafter S07) to derive a combined Hubble diagram for 69 GRBs, later updated by Cardone et al. (2009, hereafter CCD09) where also a sixth correlation between the break time $T_a$ and the luminosity at break time $L_X(T_a)$ of the X\,-\,ray afterglow \citep{DCC08,D10,D11} has been considered.

Recently, a larger GRBs catalog has been compiled by Xiao \& Schaefer (2009, hereafter XS09) giving, for each object, the values of the different quantities (if available) needed to check the five correlations above. We use this sample here as input for our analysis referring the interested reader to XS10 for details on the catalog construction and observable quantities determination.

\subsection{Bayesian fitting procedure}

Eq.(\ref{eq: loglog}) is a linear relation which can be fitted to a given dataset $(x_i, y_i)$ in order to determine the two calibration parameters
$(a, b)$. Moreover, although there is still not a theoretical model explaining any of the empirical 2D correlations in terms of GRB physics, we nevertheless expect that the wide range of GRB properties makes the objects scatter around this (unknown) idealized model. As a consequence, the above linear relations will be affected by an intrinsic scatter $\sigma_{int}$ which has to be determined together with the calibration coefficients $(a, b)$.  To this aim, in the following we will resort to a Bayesian motivated technique \cite{Dago05} thus maximizing the likelihood function ${\cal{L}}(a, b, \sigma_{int}) = \exp{[-L(a, b, \sigma_{int})]}$ with\,:

\begin{eqnarray}
L(a, b, \sigma_{int}) & = &
\frac{1}{2} \sum{\ln{(\sigma_{int}^2 + \sigma_{Y_i}^2 + a^2
\sigma_{X_i}^2)}} \nonumber \\
~ & + & \frac{1}{2} \sum{\frac{(Y_i - a X_i - b)^2}{\sigma_{int}^2 +
\sigma_{Y_i}^2 + a^2 \sigma_{X_i}^2}}
\label{eq: deflike}
\end{eqnarray}
with $(X_i, Y_i) = (\log{x_i}, \log{y_i})$ and the sum is over the ${\cal{N}}$ objects in the sample. Note that, actually, this maximization is performed in the two parameter space $(a, \sigma_{int})$ since $b$ may be estimated analytically as\,:

\begin{equation}
b = \left [ \sum{\frac{Y_i - a X_i}{\sigma_{int}^2 + \sigma_{Y_i}^2 + a^2
\sigma_{X_i}^2}} \right ] \left [\sum{\frac{1}{\sigma_{int}^2 + \sigma_{Y_i}^2 + a^2
\sigma_{X_i}^2}} \right ]^{-1}
\label{eq: calca}
\end{equation}
so that we will not consider it anymore as a fit parameter.

It is worth noting that the Bayesian approach allows to find out what is the most likely set of parameters within a given theory, but does not tell
us whether this model fits well or not the data. An easy way to quantitatively estimate the goodness of the fit is obtained considering the median and root mean square of the best fit residuals, defined as $\delta = Y_{obs} - Y_{fit}$ which we will also compute for the different 2D correlations we will consider.

The Bayesian approach used here also allows us to quantify the uncertainties on the fit parameters. To this aim, for a given parameter $p_i$, we first compute the marginalized likelihood ${\cal{L}}_i(p_i)$ by integrating over the other parameter. The median value for the parameter $p_i$ is then found by solving\,:

\begin{equation}
\int_{p_{i,min}}^{p_{i,med}}{{\cal{L}}_i(p_i) dp_i} = \frac{1}{2}
\int_{p_{i,min}}^{p_{i,max}}{{\cal{L}}_i(p_i) dp_i} \ .
\label{eq: defmaxlike}
\end{equation}
The $68\%$ ($95\%$) confidence range $(p_{i,l}, p_{i,h})$ are then found by solving\,:

\begin{equation}
\int_{p_{i,l}}^{p_{i,med}}{{\cal{L}}_i(p_i) dp_i} = \frac{1 - \varepsilon}{2}
\int_{p_{i,min}}^{p_{i,max}}{{\cal{L}}_i(p_i) dp_i} \ ,
\label{eq: defpil}
\end{equation}

\begin{equation}
\int_{p_{i,med}}^{p_{i,h}}{{\cal{L}}_i(p_i) dp_i} = \frac{1 - \varepsilon}{2}
\int_{p_{i,min}}^{p_{i,max}}{{\cal{L}}_i(p_i) dp_i} \ ,
\label{eq: defpih}
\end{equation}
with $\varepsilon = 0.68$ (0.95) for the $68\%$ ($95\%$) range respectively. Actually, in order to sample the parameter space, we use a Markov Chain Monte Carlo (MCMC) method running two parallel chains and using the Gelman\,-\,Rubin (1992) test to check convergence. The confidence ranges are then obtained considering the histograms of the parameters from the merged chain after burn in cut and thinning.

\subsection{GRBs luminosity distances}

A preliminary step in the analysis of all the 2D correlations hinted at above is the determination of the luminosity $L$ or the collimated energy $E_{\gamma}$ entering as $Y$ variable in the $Y$\,-\,$X$ scaling laws. As shown by Eqs.(\ref{eq: lpbolo})\,-\,(\ref{eq: egamma}), one has to first determine the GRBs luminosity distance over a redshift range where the linear Hubble law does not hold anymore. As such, one should estimate the luminosity distance as\,:

\begin{equation}
d_L(z) = \frac{c (1 + z)}{H_0} \int_{0}^{z}{\frac{dz'}{E(z')}}
\label{eq: dldef}
\end{equation}
with $H_0 = 100 h \ {\rm km/s/Mpc}$ the present day Hubble constant and $E(z) = H(z)/H_0$ the dimensionless Hubble parameter which depends on the adopted cosmological model thus leading to the well known {\it circularity problem} (i.e., one would like to use GRBs to probe the cosmological model, but needs the cosmological model itself to get the GRBs Hubble diagram).

Different strategies have been developed to tackle this problem. The simplest one is to assume a fiducial cosmological model and determine its parameters by fitting, e.g., the SNeIa Hubble diagram. Motivated by its agreement with a wide set of data, one usually adopt the $\Lambda$CDM model as fiducial one thus setting\,:

\begin{equation}
E^2(z) = \Omega_M (1 + z)^3 + \Omega_{\Lambda}
\label{eq: ezlcdm}
\end{equation}
with $\Omega_{\Lambda} = 1 - \Omega_M$ because of the spatial flatness assumption. In order to determine the parameters $(\Omega_M, h)$, we maximize the likelihood function ${\cal{L}}(\Omega_M, h) \propto \exp{(-\chi^2/2)}$ with\,:

\begin{eqnarray}
\chi^2(\Omega_M, h) & = & \sum_{i = 1}^{{\cal{N}}_{SNeIa}}{\left [ \frac{\mu_i^{obs} - \mu^{th}(z_i, \Omega_M, h)}{\sigma_{\mu_i}} \right ]^2}
\nonumber \\
~ & + & \left ( \frac{\omega_M^{obs} - \omega_M^{th}}{\sigma_{\omega_M}} \right )^2 + \left ( \frac{h^{obs} - h}{\sigma_h} \right )^2 \ .
\label{eq: defchilcdm}
\end{eqnarray}
As input data, we use the Union2 SNeIa sample \citep{Union2} to get $(\mu_i^{obs}, \sigma_{\mu_i})$ for ${\cal{N}}_{SNeIa} = 557$ objects over the redshift range $(0.015, 1.4)$ and set $(\omega_M^{obs}, \sigma_{\omega_M}) = (0.1356, 0.0034)$ for the matter physical density $\omega_M = \Omega_M h^2$ and $(h, \sigma_h) = (0.742, 0.036)$ for the Hubble constant in agreement with the results of the WMAP7 \citep{WMAP7} and SHOES \citep{SHOES} teams, respectively. The best fit values turn out to be $(\Omega_M, h) = (0.261, 0.722)$, while median value and confidence ranges read\,:

\begin{displaymath}
\Omega_M = 0.259_{-0.016 \ -0.028}^{+0.019 \ +0.032} \ \ , \ \ h = 0.723_{-0.025 \ -0.045}^{+0.025 \ +0.046} \ \ .
\end{displaymath}
Although the $\Lambda$CDM model fits remarkably well the data, it is nevertheless worth stressing that a different cosmological model would give different values for $d_L(z)$ and hence different values for $(L, E_{\gamma})$ thus impacting the estimate of the calibration parameters $(a, \sigma_{int})$. Although CCD09 have shown that this effect is quite small for a large range of phenomenological dark energy models\footnote{CCD09 adopted the CPL \citep{CP01,L03} parametrization of the dark energy equation of state, $w = w_0 + w_a z/(1 + z)$, and explored the dependence of $(a,    \sigma_{int})$ on $(w_0, w_a)$ for different values of $\Omega_M$. While clear trends of $(a, \sigma_{int})$ with $(\Omega_m, w_0, w_a)$ were obtained, the relative change is negligibly small. It is worth stressing, however, that the calibration parameters may change more if the adopted cosmological model is radically different from the $\Lambda$CDM one (see, e.g, Diaferio et al. 2011 for an illuminating example).}, it is nevertheless worth be conservative and look for model independent approaches. A first step towards this aim is to resort to cosmography \citep{W72,V04}, i.e., to expand the scale factor $a(t)$ to the fifth order and then consider the luminosity distance as a function of the cosmographic parameters. Indeed, such a kinematic approach only relies on the validity of assumption of the Robertson\,-\,Walker metric, while no assumption on either the cosmological model or the theory of gravity is needed since the Friedmann equations are never used. We use the expansion of $d_L(z)$ to the fifth order in $z$ found in Capozziello et al. (2008) and determine the Hubble constant $h$, the deceleration $q_0$, jerk $j_0$, snap $s_0$ and lerk $l_0$ parameters by fitting the Union2 SNeIa data with the prior on $h$ coming from the SHOES team quoted above. Note that, in this case, the likelihood function is defined as before, but the pseudo\,-\,$\chi^2$ is now given by Eq.(\ref{eq: defchilcdm}) without the term depending on $\omega_M$ since now $\Omega_M$ is no more a parameter. Using a similar MCMC algorithm (but now running three parallel chains), we find as best fit values\,:

\begin{displaymath}
(h, q_0, j_0, s_0, l_0) = (0.741, -0.56, 0.66, -0.41, 3.59) \ ,
\end{displaymath}
while median values and confidence ranges read\,:

\begin{displaymath}
h = 0.741_{-0.035 \ -0.072}^{+0.036 \ +0.071} \ \ ,
\ \ q_0 = -0.45_{-0.05 \ -0.14}^{+0.05 \ +0.10} \ \ ,
\end{displaymath}

\begin{displaymath}
j_0 = 0.00_{-0.12 \ -0.35}^{+0.13 \ +0.79} \ \ ,
\ \ s_0 = 0.29_{-0.20 \ -1.08}^{+0.74 \ +1.57} \ \ ,
\end{displaymath}

\begin{displaymath}
l_0 = -0.07_{-4.71 \ -10.86}^{+4.77 \ +12.27} \ \ ,
\end{displaymath}
in good agreement with previous results in literature \citep{Vit09,XW10,BCC10} using different datasets.

As a further step towards a fully model independent estimate of the GRBs luminosity distances, one can use SNeIa as distance indicator based on the naive observations that a GRBs at redshift $z$ must have the same distance modulus of SNeIa having the same redshift.Interpolating therefore the SNeIa Hubble diagram gives therefore the value of $\mu(z)$ for a subset of the GRBs sample with $z \le 1.4$ which can then be used to calibrate the 2D correlations \citep{K08,L08,WZ08}. Assuming that this calibration is redshift independent, one can then build up the Hubble diagram at higher redshifts using the calibrated correlations for the remaining GRBs in the sample. CCD09 have used a  similar approach based on the local regression technique \citep{C79,CD88,L99} which combines much of the simplicity of linear least squares regression with the flexibility of nonlinear regression. The basic idea relies on fitting simple models to localized subsets of the data to build up a function that describes the deterministic part of the variation in the data, point by point. Actually, one is not required to specify a global function of any form to fit a model to the data so that there is no ambiguity in the choice of the interpolating function. Indeed, at each point, a low degree polynomial is fit to a subset of the data containing only those points which are nearest to the point whose response is being estimated. The polynomial is fit using weighted least squares with a weight function which quickly decreases with the distance from the point where the model has to be recovered. We apply local regression to estimate the distance modulus $\mu(z)$ from the Union2 SNeIa sample following the steps schematically sketched in CCD09 which we refer the reader to for further details and the demonstration of the reliability of the inferred luminosity distances.

\begin{table*}
\caption{Constraints on the calibration parameters $(a, b, \sigma_{int})$ for the 2D correlations considered in the text. Columns are as follows\,: 1. correlation id, 2. number of GRBs used, 3. reduced $\chi^2$, 4. best fit parameters, 5., 6., 7. median values and $68\%$ confidence ranges for $(a, b, \sigma_{int})$, respectively. For each correlation, there are three rows referring to the results obtained using the F, C and LR distances.}
\label{tab: calpar}
\begin{center}
\begin{tabular}{|c|c|c|c|c|c|c|}
\hline Id & ${\cal{N}}$ & $\chi^2/d.o.f.$ & $(a, b, \sigma_{int})_{bf}$ & $a$ & $b$ & $\sigma_{int}$ \\
\hline \hline
~ & ~ & ~ & ~ & ~ & ~ & ~    \\

$L$\,-\,$E_{peak}$ & 39 & 1.05 & (1.05, 49.75, 0.46) & $0.99_{-0.22 \ -0.43}^{+0.21 \ +0.41}$ & $49.61_{-0.21 \ -0.97}^{+0.64 \ +1.14}$ & $0.46_{-0.05 \ -0.11}^{+0.08 \ +0.16}$    \\

~ & ~ & ~ & ~ & ~ & ~ & ~   \\

$L$\,-\,$E_{peak}$ & 39 & 1.04 & (1.06, 49.66, 0.47) & $1.02_{-0.20 \ -0.43}^{+0.21 \ +0.44}$ & $49.53_{-0.10 \ -0.71}^{+0.57 \ +1.15}$ & $0.47_{-0.05 \ -0.10}^{+0.07 \ +0.15}$       \\

~ & ~ & ~ & ~ & ~ & ~ & ~   \\

$L$\,-\,$E_{peak}$ & 39 & 1.08 & (0.95, 50.00, 0.43) & $0.84_{-0.23 \ -0.44}^{+0.23 \ +0.47}$ & $50.02_{-0.21 \ -0.72}^{+0.68 \ +1.24}$ & $0.45_{-0.07 \ -0.12}^{+0.05 \ +0.17}$       \\

~ & ~ & ~ & ~ & ~ & ~ & ~  \\

\hline

~ & ~ & ~ & ~ & ~ & ~ & ~   \\

$L$\,-\,$\tau_{lag}$ & 27 & 1.07 & (-0.70, 51.60, 0.47) & $-0.65_{-0.16 \ -0.31}^{+0.17 \ +0.36}$ & $51.60_{-0.05 \ -0.17}^{+0.12 \ +0.25}$ & $0.48_{-0.06 \ -0.11}^{+0.06 \ +0.19}$    \\

~ & ~ & ~ & ~ & ~ & ~ & ~   \\

$L$\,-\,$\tau_{lag}$ & 27 & 1.09 & (-0.68, 51.57, 0.47) & $-0.64_{-0.15 \ -0.33}^{+0.14 \ +0.28}$ & $51.60_{-0.08 \ -0.21}^{+0.07 \ +0.17}$ & $0.49_{-0.07 \ -0.13}^{+0.08 \ +0.20}$   \\

~ & ~ & ~ & ~ & ~ & ~ & ~   \\

$L$\,-\,$\tau_{lag}$ & 27 & 1.09 & (-0.67, 51.67, 0.43) & $-0.62_{-0.14 \ -0.33}^{+0.15 \ +0.27}$ & $51.70_{-0.07 \ -0.18}^{+0.10 \ +0.18}$ & $0.44_{-0.06 \ -0.12}^{+0.07 \ +0.16}$    \\

~ & ~ & ~ & ~ & ~ & ~ & ~   \\

\hline

~ & ~ & ~ & ~ & ~ & ~ & ~   \\

$L$\,-\,$\tau_{RT}$ & 31 & 1.16 & (-1.09, 51.81, 0.45) & $-1.02_{-0.19 \ -0.47}^{+0.24 \ +0.46}$ & $51.82_{-0.04 \ -0.10}^{+0.05 \ +0.12}$ & $0.46_{-0.06 \ -0.11}^{+0.08 \ +0.20}$    \\

~ & ~ & ~ & ~ & ~ & ~ & ~   \\

$L$\,-\,$\tau_{RT}$ & 31 & 1.20 & (-1.03, 51.78, 0.46) & $-0.92_{-0.24 \ -0.49}^{+0.25 \ +0.47}$ & $51.83_{-0.07 \ -0.15}^{+0.04 \ +0.10}$ & $0.48_{-0.07 \ -0.13}^{+0.08 \ +0.18}$    \\

~ & ~ & ~ & ~ & ~ & ~ & ~   \\

$L$\,-\,$\tau_{RT}$ & 31 & 1.10 & (-1.07, 51.88, 0.40) & $-0.97_{-0.22 \ -0.47}^{+0.22 \ +0.45}$ & $51.88_{-0.02 \ -0.10}^{+0.07 \ +0.14}$ & $0.41_{-0.06 \ -0.11}^{+0.07 \ +0.15}$    \\

~ & ~ & ~ & ~ & ~ & ~ & ~   \\

\hline
~ & ~ & ~ & ~ & ~ & ~ & ~   \\

$L$\,-\,$V$ & 37 & 1.10 & (0.34, 52.60, 0.61) & $0.33_{-0.20 \ -0.52}^{+0.18 \ +0.43}$ & $52.53_{-0.16 \ -0.47}^{+0.25 \ +0.56}$ & $0.62_{-0.07 \ -0.14}^{+0.10 \ +0.20}$    \\

~ & ~ & ~ & ~ & ~ & ~ & ~   \\

$L$\,-\,$V$ & 37 & 1.09 & (0.40, 52.65, 0.61) & $0.37_{-0.25 \ -0.50}^{+0.21 \ +0.49}$ & $52.66_{-0.30 \ -0.69}^{+0.16 \ +0.54}$ & $0.63_{-0.08 \ -0.14}^{+0.09 \ +0.23}$    \\

~ & ~ & ~ & ~ & ~ & ~ & ~   \\

$L$\,-\,$V$ & 37 & 1.10 & (0.37, 52.73, 0.56) & $0.33_{-0.22 \ -0.52}^{+0.19 \ +0.41}$ & $52.64_{-0.22 \ -0.61}^{+0.24 \ +0.52}$ & $0.57_{-0.07 \ -0.13}^{+0.09 \ +0.13}$    \\

~ & ~ & ~ & ~ & ~ & ~ & ~   \\

\hline

~ & ~ & ~ & ~ & ~ & ~ & ~   \\

$E_{\gamma}$\,-\,$E_{peak}$ & 14 & 1.17 & (1.77, 46.71, 0.17) & $1.71_{-0.24 \ -0.65}^{+0.22 \ +0.44}$ & $47.01_{-0.58 \ -1.09}^{+0.23 \ +1.01}$ & $0.20_{-0.06 \ -0.10}^{+0.07 \ +0.19}$     \\

~ & ~ & ~ & ~ & ~ & ~ & ~   \\

$E_{\gamma}$\,-\,$E_{peak}$ & 14 & 1.21 & (1.71, 46.80, 0.18) & $1.62_{-0.29 \ -0.64}^{+0.24 \ +0.51}$ & $47.28_{-0.71 \ -1.35}^{+0.27 \ +1.10}$ & $0.21_{-0.06 \ -0.11}^{+0.07 \ +0.16}$    \\

~ & ~ & ~ & ~ & ~ & ~ & ~   \\

$E_{\gamma}$\,-\,$E_{peak}$ & 14 & 1.16 & (1.72, 46.81, 0.17) & $1.66_{-0.28 \ -0.57}^{+0.26 \ +0.55}$ & $47.16_{-0.66 \ -1.31}^{+0.38 \ +0.95}$ & $0.20_{-0.07 \ -0.14}^{+0.07 \ +0.18}$    \\

~ & ~ & ~ & ~ & ~ & ~ & ~   \\

\hline

~ & ~ & ~ & ~ & ~ & ~ & ~   \\

$E_{iso}$\,-\,$E_{peak}$ & 40 & 0.79 & (1.58, 49.16, 0.52) & $1.54_{-0.26 \ -0.50}^{+0.27 \ +0.55}$ & $49.26_{-0.45 \ -1.04}^{+0.45 \ + 1.03}$ & $0.51_{-0.07 \ -0.13}^{+0.09 \ +0.21}$    \\

~ & ~ & ~ & ~ & ~ & ~ & ~   \\

$E_{iso}$\,-\,$E_{peak}$ & 40 & 0.81 & (1.24, 49.57, 0.52) & $1.00_{-0.33 \ -0.62}^{+0.37 \ +0.75}$ & $50.16_{-0.30 \ -1.27}^{+0.99 \ + 1.69}$ & $0.52_{-0.08 \ -0.16}^{+0.11 \ +0.24}$    \\

~ & ~ & ~ & ~ & ~ & ~ & ~   \\

$E_{iso}$\,-\,$E_{peak}$ & 40 & 0.79 & (1.22, 50.03, 0.51) & $1.04_{-0.38 \ -0.70}^{+0.35 \ +0.73}$ & $50.09_{-0.27 \ -1.26}^{+0.97 \ + 1.77}$ & $0.52_{-0.08 \ -0.15}^{+0.12 \ +0.27}$    \\

~ & ~ & ~ & ~ & ~ & ~ & ~   \\

\hline
\end{tabular}
\end{center}
\end{table*}

\section{Calibration parameters}

While the $X$ quantities are directly observed for each GRBs, the determination of $Y$ (either the luminosity $L$ or the collimated energy $E_{\gamma}$) needs for the object's luminosity distance. The three methods described above allows us to get three different values for $Y$ so that it is worth investigating whether this has any significant impact on the calibration parameters $(a, b, \sigma_{int})$ for the correlations of interest. We will refer hereafter to the three samples with the $Y$ quantities estimated using the luminosity distance from the fiducial $\Lambda$CDM cosmological model, the cosmographic parameters and the local regression method as the $F$, $C$ and $LR$ samples, respectively.

As a preliminary caveat, it is worth stressing that the error on $Y$ comes from two different contributions. First, there is the uncertainty obtained by propagating those on the observed quantities (i.e., the flux $P$ and the Band parameters for $L$ and the fluence $S$, the jet angle $\theta_{jet}$ and again the Band parameters for $E_{\gamma}$) which we assume to be uncorrelated. Second, there is the uncertainty on $d_L(z)$ which we estimate from the MCMC chain for the $F$ and $C$ cases, while it is output from the method for the $LR$ case. It is worth noting that, given the large SNeIa sample used in the fit of $\Lambda$CDM and cosmographic parameters and in the local regression technique, the error on the luminosity distance is actually quite small and always smaller than the one from the measurement uncertainties. It is this latter term that one should minimize in order to get better determined $L$ and/or $E_{\gamma}$ values and hence put stronger constraints on the slope, zeropoint and intrinsic scatter.

In order to constrain the calibration parameters $(a, b, \sigma_{int})$, we can use the Bayesian procedure described above with the three GRBs samples as input to the likelihood analysis. However, the $F$, $C$ and $LR$ samples can not contain the same number of objects since the GRBs luminosity distance (and hence the $Y$ quantities) can be estimated only for $z_{min} \le z \le z_{max}$ with $(z_{min}, z_{max})$ depending on the adopted method. If we rely on the fiducial $\Lambda$CDM model, we can predict $d_L(z)$ at every $z$ so that $(z_{min}, z_{max}) = (0, \infty)$ with the only caveat that one is implicitly assuming that a model fitted over the range $(0.015, 1.4)$ probed by the Union2 SNeIa sample can be extrapolated to the full evolutionary history of the universe. The $C$ sample is based on the use of the cosmographic parameters fitted to the data having checked that the fifth order expansion of $d_L(z)$ is reliable over the Union2 SNeIa redshift range. However, the larger is $z$, the higher is the order one has to include in the $d_L(z)$ Taylor series to get accurate results so that one must set $(z_{min}, z_{max}) = (0, 1.4)$ in order to not bias the determination of the $Y$ quantities because of an inaccurate distance approximation. Finally, the local regression method can only be applied to estimate $d_L(z)$ over the redshift range probed by the data used for the interpolation so that one get the constraint $(z_{min}, z_{max}) = (0.015 , 1.4)$. In order to homogenize the three samples, we therefore fit the 2D correlations using only $F$, $C$ and $LR$ GRBs with $z \le 1.4$ so that we can meaningfully compare the results for the three distance estimate methods.

Table\,\ref{tab: calpar} summarizes the constraints on the calibration parameters for the three different samples. Note that we have also included the $E_{iso}$\,-\,$E_{peak}$ correlation \citep{A09} with $E_{iso} = E_{\gamma}/F_{beam}$ since, although refers to the same quantities used in the $E_{\gamma}$\,-\,$E_{peak}$ correlation, it is based on a larger number of GRBs (given that one does not need a measurement of the jet opening angle $\theta_{jet}$). As a general result, we find that the fit is always quite good with reduced $\chi^2$ values close to $1$ in all the cases independently on the 2D correlation considered and the distance estimate method adopted. As such, the results from the different sample are statistically equivalent and can be safely compared.

Although the $68$ and $95\%$ confidence ranges for the calibration parameters $(a, b, \sigma_{int})$ for a given correlation overlap quite well for the $F$, $C$ and $LR$ samples, the best fit coefficients and the median values clearly show that the calibration based on the fiducial $\Lambda$CDM model leads to steeper scaling laws for most of the cases. On the contrary, shallower slopes are obtained using the $C$ or $LR$ samples with the $L$\,-\,$V$ relation as unique exception\footnote{Actually, the $L$\,-\,$V$ relation is the shallowest one and has the largest intrinsic scatter among the six 2D correlations investigated in this paper so that a different trend of the slope with the sample could also be a statistical artifact.}. Although the differences in the slopes are not statistically meaningful because of the large uncertainties, it is nevertheless worth investigating whether such an effect can be ascribed to the distance determination. To this end, for a given GRB and $Y$\,-\,$X$ correlation, let us define the following quantities\,:

\begin{displaymath}
\Delta_X Y = (a_1 X + b_1) - (a_2 X + b_2) = (a_1 - a_2) X + (b_1 - b_2)  \ ,
\end{displaymath}

\begin{displaymath}
\Delta_{d_L} Y = 2 \log{[d_{L1}(z)/d_{L2}(z)]}  \ .
\end{displaymath}
While $\Delta_{X} Y$ gives the difference in the $Y$ values predicted using the correlations with coefficients $(a_1, b_1)$ and $(a_2, b_2)$, $\Delta_{d_L} Y$ quantifies the effect of estimating the model dependent quantity $Y$ using two different luminosity distances. Should the change in the slope be the outcome of compensating the offset due to different luminosity distances assumptions, one should get $\Delta_{X} Y \simeq \Delta_{d_L} Y$. Actually, this is not the case. Considering, e.g, the $L$\,-\,$E_{peak}$ correlation, a weighted average gives\,:

\begin{displaymath}
\langle \Delta_{X} Y \rangle = 0.19 \neq  \langle \Delta_{d_L} Y  \rangle = 0.09
\end{displaymath}
using the $F$ and $C$ samples and

\begin{displaymath}
\langle \Delta_{X} Y \rangle = -0.25 \neq  \langle \Delta_{d_L} Y  \rangle = -0.02
\end{displaymath}
for the $F$ vs $LR$ samples. We therefore find that the change in the slope is not induced by the different luminosity distances adopted. Actually, the use of different samples has also an impact on the intrinsic scatter determination with the $LR$ sample leading to smaller best fit and median $\sigma_{int}$ values for all the six correlations considered. Since $a$ and $\sigma_{int}$ correlate, the change in the slope is not only due to the change in the luminosity distances, but also of the intrinsic scatter. Discriminating among these two effects and quantifying their respective contribution is not possible so that we can not draw any definitive conclusion on the impact of the calibration method on the slope of the 2D scaling relations.

\begin{table*}
\caption{Constraints on the calibration parameters $(a, b, \alpha, \sigma_{int})$ for the 2D correlations considered in the text and fitted to the $F$ sample. Columns are as follows\,: 1. correlation id, 2. reduced $\chi^2$, 3. best fit parameters, 4., 5., 6., 7. median values and $68\%$ confidence ranges for the fit parameters. The number of GRBs used in the same as in Table\,\ref{tab: calpar}.}
\label{tab: plpar}
\begin{center}
\begin{tabular}{|c|c|c|c|c|c|c|}
\hline
Id & $\chi^2/d.o.f.$ & $(a, b, \alpha, \sigma_{int})_{bf}$ & $a$ & $b$ & $\alpha$ & $\sigma_{int}$ \\
\hline \hline

~ & ~ & ~ & ~ & ~ & ~ & ~    \\

$L$\,-\,$E_{peak}$ & 1.08 & (0.89, 50.77, 0.67, 0.41) & $0.82_{-0.19}^{+0.20}$ & $50.75_{-0.65}^{+0.59}$ & $0.68_{-0.30}^{+0.18}$ & $0.44_{-0.06}^{+0.06}$ \\

~ & ~ & ~ & ~ & ~ & ~ & ~    \\

$L$\,-\,$\tau_{lag}$ & 1.10 & (-0.58, 50.38, 0.88, 0.39) & $-0.57_{-0.14}^{+0.12}$ & $50.39_{-0.58}^{+0.66}$ & $0.87_{-0.33}^{+0.30}$ & $0.41_{-0.06}^{+0.07}$ \\

~ & ~ & ~ & ~ & ~ & ~ & ~    \\

$L$\,-\,$\tau_{RT}$ & 1.16 & (-0.79, 51.22, 0.65, 0.44) & $-0.64_{-0.25}^{+0.21}$ & $50.97_{-0.75}^{+0.70}$ & $0.76_{-0.33}^{+0.36}$ & $0.47_{-0.06}^{+0.07}$ \\

~ & ~ & ~ & ~ & ~ & ~ & ~    \\

$L$\,-\,$V$ & 1.23 & (-0.10, 50.27, 1.07, 0.54) & $-0.05_{-0.37}^{+0.26}$ & $50.29_{-0.96}^{+0.91}$ & $1.04_{-0.50}^{+0.52}$ & $0.57_{-0.08}^{+0.11}$ \\

~ & ~ & ~ & ~ & ~ & ~ & ~    \\

$E_{\gamma}$\,-\,$E_{peak}$ & 1.37 & (1.57, 50.64, 0.00, 0.14) & $1.48_{-0.26}^{+0.23}$ & $50.42_{-0.57}^{+0.51}$ & $0.13_{-0.28}^{+0.31}$ & $0.20_{-0.06}^{+0.07}$ \\

~ & ~ & ~ & ~ & ~ & ~ & ~    \\

$E_{iso}$\,-\,$E_{peak}$ & 0.81 & (1.39, 51.58, 0.58, 0.48) & $1.29_{-0.28}^{+0.28}$ & $51.57_{-0.81}^{+0.84}$ & $0.58_{-0.45}^{+0.44}$ & $0.50_{-0.08}^{+0.10}$ \\

~ & ~ & ~ & ~ & ~ & ~ & ~    \\
\hline
\end{tabular}
\end{center}
\end{table*}

\begin{table*}
\caption{Constraints on the calibration parameters $(a_0, \log{a_1}, b_0, \log{b_1}, \sigma_{int})$ for the 2D correlations considered in the text and fitted to the $F$ sample. Columns are as follows\,: 1. correlation id, 2. reduced $\chi^2$, 3. best fit parameters, 4., 5., 6., 7., 8. median values and $68\%$ confidence ranges for the fit parameters. The number of GRBs used in the same as in Table\,\ref{tab: calpar}.}
\label{tab: linpar}
\begin{center}
\begin{tabular}{|c|c|c|c|c|c|c|c|}
\hline
Id & $\chi^2/d.o.f.$ & $(a_0, \log{a_1}, b_0, \log{b_1}, \sigma_{int})_{bf}$ & $a_0$ & $\log{a_1}$ & $b_0$ & $\log{b_1}$ & $\sigma_{int}$ \\
\hline \hline

~ & ~ & ~ & ~ & ~ & ~ & ~ & ~    \\

$L$\,-\,$E_{peak}$ & 1.12 & (0.89, -2.69, 51.49, -0.21, 0.41) & $0.89_{-0.25}^{+0.22}$ & $-3.05_{-3.63}^{+2.13}$ & $51.87_{-0.46}^{+0.19}$ & $-0.78_{-3.06}^{+0.64}$ & $0.44_{-0.05}^{+0.03}$ \\

~ & ~ & ~ & ~ & ~ & ~ & ~ & ~    \\

$L$\,-\,$\tau_{lag}$ & 1.13 & (-0.60, -4.43, 51.31, -0.07, 0.40) & $-0.64_{-0.19}^{+0.16}$ & $-2.78_{-2.85}^{+1.61}$ & $51.61_{-0.45}^{+0.49}$ & $-0.28_{-2.83}^{+0.27}$ & $0.43_{-0.06}^{+0.08}$ \\

~ & ~ & ~ & ~ & ~ & ~ & ~ & ~    \\

$L$\,-\,$\tau_{RT}$ & 1.17 & (-1.56, -0.08, 52.38, -0.70, 0.44) & $-0.95_{-0.46}^{+0.33}$ & $-1.81_{-3.90}^{+1.59}$ & $52.42_{-0.53}^{+0.19}$ & $-1.81_{-3.91}^{+1.60}$ & $0.47_{-0.06}^{+0.07}$ \\

~ & ~ & ~ & ~ & ~ & ~ & ~ & ~    \\

$L$\,-\,$V$ & 1.36 & (-0.15, -2.16, 51.40, 0.03, 0.52) & $-0.10_{-0.44}^{+0.41}$ & $-1.79_{-4.81}^{+1.49}$ & $51.85_{-0.53}^{+0.39}$ & $-0.77_{-2.86}^{+0.76}$ & $0.58_{-0.08}^{+0.10}$ \\

~ & ~ & ~ & ~ & ~ & ~ & ~ & ~    \\

$E_{\gamma}$\,-\,$E_{peak}$ & 1.70 & (0.79, 0.04, 50.37, -0.56, 0.10) & $1.40_{-0.47}^{+0.23}$ & $-1.70_{-2.67}^{+1.63}$ & $50.61_{-0.15}^{+0.08}$ & $-2.86_{-3.84}^{+1.91}$ & $0.17_{-0.06}^{+0.08}$ \\

~ & ~ & ~ & ~ & ~ & ~ & ~ & ~    \\

$E_{iso}$\,-\,$E_{peak}$ & 0.79 & (0.94, -0.23, 52.00, -0.15, 0.49) & $1.27_{-0.50}^{+0.33}$ & $-1.78_{-2.54}^{+1.61}$ & $52.56_{-0.34}^{+0.16}$ & $-1.73_{-2.41}^{+1.42}$ & $0.50_{-0.08}^{+0.10}$ \\

~ & ~ & ~ & ~ & ~ & ~ & ~ & ~    \\
\hline
\end{tabular}
\end{center}
\end{table*}

\section{Evolution with redshift}

As more and more data add to the available GRBs dataset, the observational evidences for the six 2D correlations we are considering become more and more reliable. On the contrary, a clear theoretical motivation is still lacking in many cases. As a consequence, it is also not clear whether the calibration parameters $(a, b, \sigma_{int})$ evolve with the redshift or not. To investigate this issue, we consider two different possibilities for the evolution with $z$. First, we consider the possibility that the slope is constant, but the zeropoint is evolving. In particular, we assume\,:

\begin{equation}
y = B (1 + z)^{\alpha} x^A \ \longrightarrow \ Y = \alpha \log{(1 + z)} + a X + b
\label{eq: plfit}
\end{equation}
with $(X, Y) = (\log{x}, \log{y})$ and $(a, b) = (A, \log{b})$. Table\,\ref{tab: plpar} summarizes the constraints on the $(a, b, \alpha, \sigma_{int})$ parameters obtained fitting Eq.(\ref{eq: plfit}) to the $z \le 1.4$ GRBs with distances estimated from the fiducial $\Lambda$CDM model. The constraints from the fit to the $C$ and $LR$ samples are consistent within the $68\%$ confidence ranges so that they are reported in Appendix for completeness, but not discussed anymore here.

Comparing to the constraints in Table\,\ref{tab: calpar} highlights some interesting lessons. First, we note that both the best fit and median values of the slope parameter $a$ are significantly shallower than in the no evolution case. However, the $68\%$ confidence ranges typically overlap quite well so that, from a statistical point of view, such a result should not be overrated. Actually, adding one more parameter introduces a degeneracy between $a$ and $\alpha$ so that one can make the best fit $Y$\,-\,$X$ relation shallower compensating the difference in the term $a X$ with the contribute from $\alpha (1 + z)$. Indeed, $\alpha$ is typically quite large the only exception being the $E_{\gamma}$\,-\,$E_{peak}$ correlation which is also the only one with the same best fit slope for the fit with and without the evolution term.  Quite surprisingly, the no evolution result (i.e., $\alpha = 0$) is consistent with the $68\%$ confidence range only for the $E_{\gamma}$\,-\,$E_{peak}$ correlation thus arguing in favor of an evolution of the calibration parameters with $z$. On the other hand, the reduced $\chi^2$ values are never smaller than those obtained for the fits with no evolution. Moreover, the intrinsic scatter is almost the same for both cases so that allowing for a  redshift evolution does not lead to statistically preferred results. As such, we consider a most conservative option to assume that the GRBs scaling relations explored here do not evolve with $z$.

Actually, such a conclusion is model dependent. As an alternative parametrization, we therefore allow for an evolution of the slope and not only the zeropoint of the 2D correlations. We fit the data using\,:

\begin{equation}
Y = (a_0 + a_1 z) X + (b_0 + b_1 z) \ ,
\label{eq: linfit}
\end{equation}
i.e., we are Taylor expanding to first order the unknown dependence of the slope and zeropoint on the redshift. Note that, actually, we should limit the redshift range to very low $z$ to have a meaningful expansion, but we extrapolate this linear relation to any $z$ as a first order guess avoiding to add further unknown fitting parameters. Morevoer, since we expect $(a_1, b_1)$ are quite small, we skip to logarithmic units for these quantities in order to be more sensitive to the tiny values. Table\,\ref{tab: linpar} summarizes the results for the fit to the $F$ sample, while a qualitatively similar conclusions can be drawn from the those to the $C$ and $LR$ samples as can be inferred from Table\,\ref{tab: linparall} in Appendix A.

As a general result, we find that the best fit parameters and the median values of the evolutionary coefficients $(\log{a_1}, \log{b_1})$ are typically quite small indicating that the dependence of both the slope and the zeropoint on the redshift is quite weak, if present at all\footnote{Note that, since we are using logarithmic units, the no evolution case can not be exactly achieved corresponding to $\log{a_1}$ and $\log{b_1}$ going to $- \infty$. However, needless to say, obtaining, e.g., $\log{a_1} \simeq -2$ is the same as stating that there is no evolution at all. Note also that we have implicitly assumed that both $a_1$ and $b_1$ are positive. We have, however, checked that the qualitative conclusions are not affected by this assumption.} . On the other hand, the slope and zeropoint at $z = 0$ are consistent within the $68\%$ confidence range with the corresponding quantities in the no evolution case. Moreover, the reduced $\chi^2$ values are typically larger than in the no evolution case so that this latter assumption is statistically preferred.

\begin{figure*}
\centering
\subfigure{\includegraphics[width=5cm]{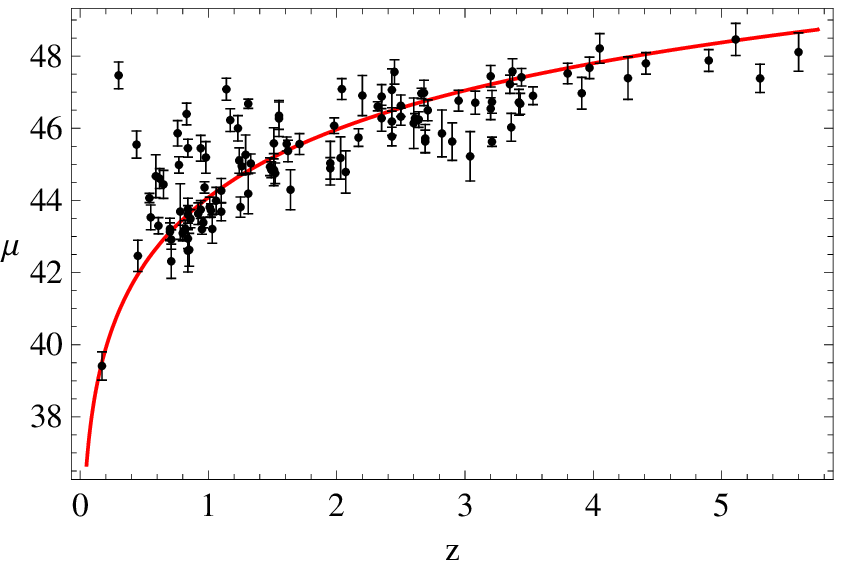}} \goodgap
\subfigure{\includegraphics[width=5cm]{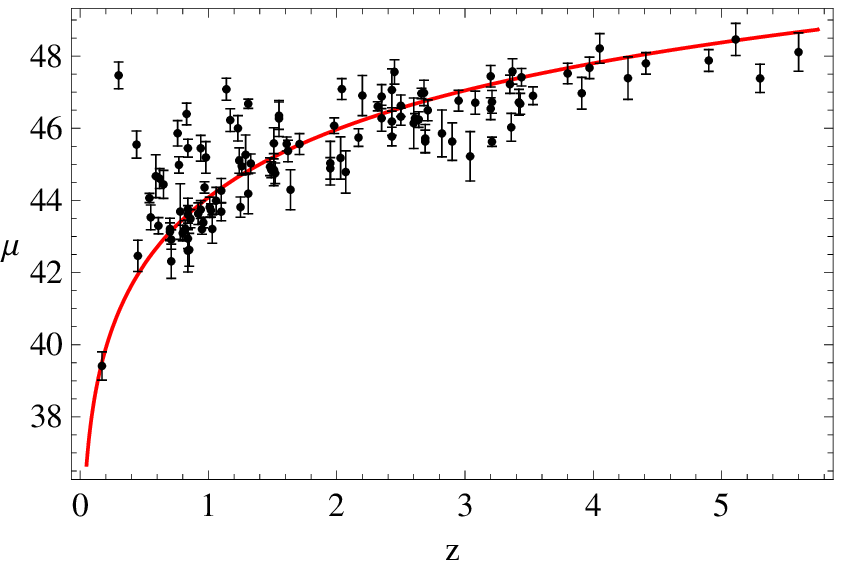}} \goodgap
\subfigure{\includegraphics[width=5cm]{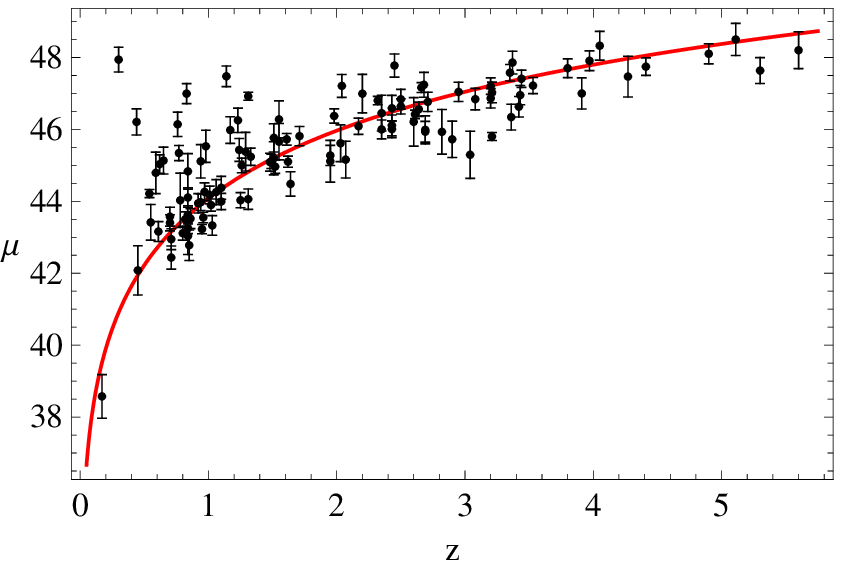}} \goodgap
\caption{GRBs Hubble diagrams (HDs) averaging over the six 2D correlations and excluding outliers (see text for the definition of outliers). Three panels refer to the HDs derived using the calibration based on the fiducial $\Lambda$CDM model (left), the cosmographic parameters (centre), and the local regression (right).}
\label{fig: grbhds}
\end{figure*}

\section{GRBs Hubble diagram}

Once the calibration parameters for a given $Y$\,-\,$X$ correlation have been obtained, it is then possible to estimate the distance modulus of a given GRB from the measured value of $X$. Indeed, for a given $Y$, the luminosity distance is\,:

\begin{displaymath}
d_L^2(z) = y/\kappa
 \end{displaymath}
 with $\kappa = 4 \pi P_{bolo}$, $\kappa = 4 \pi  S_{bolo} F_{beam}/(1 + z)$  and $\kappa = 4 \pi  S_{bolo}/(1 + z)$ for $Y = L$, $Y = E_{\gamma}$ and $E_{iso}$, respectively. Using the definition of distance modulus and estimating $Y$ from $X$ through the $Y$\,-\,$X$ correlation, we then get\,:

 \begin{eqnarray}
 \mu(z) & = & 25 + 5 \log{d_L(z)} \nonumber \\
~ & = & 25 + (5/2) \log{(y/\kappa)} \nonumber \\
~ & = & 25 + (5/2) (a X + b - \log{\kappa})
\label{eq: muval}
\end{eqnarray}
where $(a, b)$ are the best fit coefficients for the given $Y$\,-\,$X$ correlation. Note that we will refer here only to the fits with no redshift evolution of the calibration parameters since the analysis in the previous section has not shown any clear evidence for a redshift dependence of $(a, b)$. Although such a result is model dependent (since we have considered only two possible evolution parameterization) and actually limited only to the redshift range $0.015 \le z \le 1.4$, we are nevertheless confident that the possible bias induced by the no evolution assumption is smaller than the present day measurement uncertainties. However, such an issue worths to be reconsidered when a larger and more precise sample will be available.

Eq.(\ref{eq: muval}) allows to compute the central value of the distance modulus relying on the measured values of the GRBs observables, i.e., the ones entering the quantity $\kappa$, and the best fit coefficients $(a, b)$ of the used correlation. However, both $\kappa$ and $(a, b)$ are known within their own uncertainties which have to be propagated to get the error on $\mu(z)$. Moreover, each correlation is affected by an intrinsic scatter which has to be taken into account into the total error budget. To this end, we adopt the procedure schematically sketched below.

\begin{enumerate}

\item{Fix $(a, \sigma_{int})$ to the $i$\,-\,th value of the Markov Chain and estimate $b$ from Eq.(\ref{eq: calca}).}

\item{Set the quantities needed to compute $X$ and hence $\kappa$ randomly sampling from Gaussian distributions centred on the measured observed values and with variance equal to the measurement error.}

\item{Compute $\mu$ from Eq.(\ref{eq: muval}) for all the 2000 $X$ values generated above and estimate the mean of the distribution.}

\item{Repeat steps $(ii)$ and $(iii)$ for all the points along the chain and estimate the statistical error on $\mu$ by symmetrizing the $68\%$ confidence range of the distribution thus obtained.}

\item{Add in quadrature the statistical error and the intrinsic scatter  $\sigma_{int}$ of the correlation to finally get the total uncertainty on the distance modulus $\mu$.}

\end{enumerate}
Eq.(\ref{eq: muval}) and the above procedure allows us to build up the Hubble diagram for all the GRBs with measured values of the quantities entering a given $Y$\,-\,$X$ correlation. It is then possible to both reduce the uncertainties and (partially) wash out the hidden systematic errors by averaging over the different correlations available for a given GRB.  Following Schaefer (2007), we finally estimate the distance modulus for the $i$\,-\,th GRB in the sample at redshift $z_i$ as\,:

\begin{equation}
\mu(z_i) = \left [ \sum_{j}{\mu_j(z_i)/\sigma_{\mu_j}^2} \right ]
\ {\times} \ \left [ \sum_{j}{1/\sigma_{\mu_j}^2} \right ]^{-1}
\label{eq: muendval}
\end{equation}
with the uncertainty given by\,:

\begin{equation}
\sigma_{\mu} = \left [ \sum_{j}{1/\sigma_{\mu_j}^2} \right ]^{-1}
\label{eq: muenderr}
\end{equation}
where the sum runs over the 2D empirical laws which can be used for the GRB considered.

As summarized in Table \ref{tab: calpar} and discussed in Sect.\,3, the best fit calibration parameters depend on the method used to set the GRBs luminosity distances. Although such a dependence is not statistically meaningful being the confidence ranges for the different cases well overlapped, it is nevertheless worth investigating whether they have an impact on the derived Hubble diagram. Moreover, averaging over more than one correlation implicitly assumes that all of them are physically motivated (i.e., they are the outcome of an unknown underlying theoretical mechanism) and not the artifact of some not well understood selection effect. It is therefore also important to check what should be the effect of a possible misleading assumption. Both these issues will be discussed in the following.

\subsection{Impact of the calibration method}

Fig.\,\ref{fig: grbhds} shows the GRBs Hubble diagrams (hereafter, HDs) obtained averaging over the six 2D correlations and using the three different calibration methods. As a guidance for the eye, the red solid line is the expected $\mu(z)$ curve for the fiducial $\Lambda$CDM model. Note that we have excluded GRBs which deviate from this line more than $1 \sigma_{\Lambda}$ with $\sigma_{\Lambda}$ the rms of $\mu_{fid} - \mu(z)$ for the full sample. Such a criterium is actually quite loose excluding only two objects out of 114 so that we are confident that no bias is induced by this cut.

As a general remark, we find that, notwithstanding the calibration method adopted, the GRBs HDs reasonably follow the $\Lambda$CDM curve although with a non negligible scatter. Quite surprisingly, the scatter is significantly larger in the range $0.4 \le z \le 1.4$ because of a set of GRBs with $\mu(z)$ lying systematically above the $\Lambda$CDM prediction. One should argue for a failure of the theoretical model, but there are actually a set of points which are hard to reconcile with any reasonable dark energy model. While this could be dropped out from the sample by an iterative selection procedure, we prefer to err on the conservative side in order to introduce any bias induced by an a priori choice of a dark energy model. It is nevertheless likely that these GRBs are outliers of one or more of the 2D scaling laws used to infer their distance modulus. Looking in detail to their $\mu(z)$ estimates, we find that the estimates obtained from the different correlations entering the averaging procedure are quite different, while this is not for the GRBs less deviating from the red line. Such a naive observation makes us argue in favor of a problem with the measurement of the quantities entering the correlations of interest or of a deviation from one of the best fit scaling relations. Investigating in detail this issue on a case\,-\,by\,-\,case basis would need to retrieve the original data for each GRB which is outside the aim of our work.

In order to compare the HDs from the three different calibration methods, we consider the values of $\Delta \mu = \mu_{fid}(z) - \mu(z)$ with $\mu_{fid}(z)$ the theoretically predicted distance modulus for the fiducial $\Lambda$CDM model. We find\,:

\begin{displaymath}
\langle \Delta \mu \rangle = -0.15 \ \ , \ \ (\Delta \mu)_{med} = 0.07 \ \ , \ \ (\Delta \mu)_{rms} = 1.15 \ \ ,
\end{displaymath}

\begin{displaymath}
\langle \Delta \mu \rangle = -0.10 \ \ , \ \ (\Delta \mu)_{med} = 0.16 \ \ , \ \ (\Delta \mu)_{rms} = 1.15 \ \ ,
\end{displaymath}

\begin{displaymath}
\langle \Delta \mu \rangle = -0.24 \ \ , \ \ (\Delta \mu)_{med} = -0.01 \ \ , \ \ (\Delta \mu)_{rms} = 1.21 \ \ ,
\end{displaymath}
for the $F$, $C$ and $LR$ samples, respectively. One could naively be surprised that the $F$ sample (whose calibration is based on the same fiducial $\Lambda$CDM model used as reference here) does not give systematically smaller deviations. Actually, the calibration is made using only GRBs with $z \le 1.4$, but the full sample is dominated by higher $z$ objects so that the mean and median of $\Delta \mu$ are not biased by the choice of the reference model. As a consequence, we find the rms $\Delta \mu$ is roughly the same for the three samples, although the mean values suggest that the calibration based on the linear regression leads to $\mu$ values that are preferentially larger than expected. From a different point of view, higher $\mu$ values at high $z$ argue in favor of a model where the transition from deceleration to acceleration takes place to a larger $z$, which can be achieved (for a dark energy model with constant equation of state) pushing $w_0$ to values smaller than -1 (i.e., going to the phantom regime) or decreasing the matter content. However, such a naive argument should be reconsidered taking care of the uncertainties on the single GRBs distance modulus estimates.

Actually, from the point of view of the impact of calibration methods, we are not interested in minimizing the mean or median values of $\Delta \mu$, but rather on comparing the $\Delta \mu$ distributions among them. Should these distributions be equal, then one could conclude that the HDs based on different calibration methods are consistent with each other. The different mean and median values seem to suggest that this is not the case. However, these distributions are quite asymmetric and wide enough to allow for a considerable overlap among them. Moreover, for each single GRB, the values of $\Delta \mu$ are consistent with each other within the uncertainties. We can therefore conclude that the HDs obtained using different calibration methods are consistent with each other within the uncertainties.

We, nevertheless, warn the reader that such a result is partially due to the large error bars. Should these be reduced, the consistency among the three different HDs could also gone lost. To this end, however, one should first reduce the intrinsic scatter of the correlations used in the average procedure since it is this term which typically dominates the error budget. At the present time, one should give off the correlations with the larger $\sigma_{int}$ values (e.g, the $L$\,-\,$V$ one), but this would lower the number of objects in the sample and increase the error because of averaging over a lower number of $\mu$ estimates. Larger GRBs samples with measured redshift and observational parameters are then needed to address the issue of the impact of calibration methods on a safer basis.

\begin{table}
\caption{Deviations from the fiducial $\Lambda$CDM distance modulus estimates for the different GRBs 2D correlations used adopting the calibration method based on the fiducial luminosity distances (i.e., the $F$ sample). We report the number of GRBs and the mean, median and rms values.}
\label{tab: dmutab}
\begin{center}
\begin{tabular}{|c|c|c|c|c|}
\hline Id & ${\cal{N}}$ & $\langle \Delta \mu \rangle$ & $(\Delta \mu)_{med}$ & $(\Delta \mu)_{rms}$ \\
\hline \hline
~ & ~ & ~ & ~ & ~     \\

$L$\,-\,$E_{peak}$ & 115 & -0.20 & 0.01 & 1.32 \\

~ & ~ & ~ & ~ & ~  \\

$L$\,-\,$\tau_{lag}$ & 60 & 0.03 & 0.09 & 0.61 \\

~ & ~ & ~ & ~ & ~  \\

$L$\,-\,$\tau_{RT}$ & 79 & -0.12 & 0.07 & 0.94 \\

~ & ~ & ~ & ~ & ~  \\

$L$\,-\,$V$ & 115 & -0.50 & 0.05 & 1.86 \\

~ & ~ & ~ & ~ & ~  \\

$E_{\gamma}$\,-\,$E_{peak}$ & 30 & 0.04 & 0.28 & 0.97 \\

~ & ~ & ~ & ~ & ~  \\

$E_{iso}$\,-\,$E_{peak}$ & 100 & -0.01 & 0.05 & 1.01 \\

~ & ~ & ~ & ~ & ~  \\

\hline
\end{tabular}
\end{center}
\end{table}

\subsection{Impact of the averaging procedure}

As yet stated above, averaging the $\mu$ values from different correlations helps reducing the total uncertainties and partially washes out the systematics connected to each single scaling relations. However, this does not come with no price to pay. Indeed, such a strategy implicitly assumes that all the scaling relations are physical ones and not the artifact of a not well controlled selection effect. This danger is actually hard to avoid since no clearly defined and well accepted theoretical model is presently available to describe the physics of GRBs and predict scaling relations in agreement with the empirically motivated ones.

As a first check, we compare the $\Delta \mu$ values obtained estimating $\mu$ using each single correlation. For the $F$ sample, we get the results summarized in Table\,\ref{tab: dmutab} where we also give the number of GRBs used. The conclusions are unaltered for the $C$ and $LR$ samples so that we do not discuss them.

While the median values of $\Delta \mu$ are roughly comparable, both $\langle \Delta \mu \rangle$ and $(\Delta \mu)_{rms}$ are definitely larger for the $L$\,-\,$E_{peak}$ and $L$\,-\,$V$ correlations. It is noteworthy that these are also the only two correlations where all the GRBs can be used to infer the HD so that one could ascribe the larger $(\Delta \mu)_{rms}$ to the inclusion of a higher number of outliers. On the other hand, this same effect can be a statistical artifact due to the use of a smaller GRBs sample. It is actually hard to understand whether a selection effect is at work here. Indeed, the inclusion of a GRB in a sample used for a given correlation depends mainly on observational requirements (e.g., the GRBs afterglow should last enough to measure the rise time) so that one could argue that this has nothing to do with its departure from a theoretical quantity such as the fiducial distance modulus. On the other hand, should selection effects induce a fake correlation, one can imagine that the deviations from the fiducial curve are randomly distributed leading to small $\langle \Delta \mu \rangle$.

Pending the questio of which relation is physical, we can quantify the impact of an incorrect assumption by evaluating again the distance moduli excluding the $L$\,-\,$V$ and $L$\,-\,$E_{peak}$ correlations. Since these are the ones with the greatest rms $\Delta \mu$ and the largest intrinsic scatter (see Table\,\ref{tab: calpar}), we expect them to have the greatest impact on the departure from the fiducial $\Lambda$CDM theoretically predicted HD. We find\,:

\begin{displaymath}
\langle \Delta \mu \rangle = -0.11 \ \ , \ \ (\Delta \mu)_{med} = 0.04 \ \ , \ \ (\Delta \mu)_{rms} = 1.05 \ \ ,
\end{displaymath}
for 106 GRBs, being the number smaller than before because some GRBs with $\mu$ determined from the $L$\,-\,$E_{peak}$ and/or $L$\,-\,$V$ correlations only now drop out of the final sample. Compared with the case with all correlations included, the mean, median and rms $\Delta \mu$ are indeed smaller. However, since we average now on a lower number of correlations, the uncertainty on $\mu$ for each GRB is now larger so that the $\Delta \mu$ values with and without these two correlations are consistent within the "$1 \sigma$ confidence ranges.

We must therefore conclude that, given the available GRBs dataset, including or not a given correlation in the averaging procedure should be a compromise between the need to avoid an uncertain systematic bias and the desire to ameliorate both statistics and precision.

\subsection{Satellite dependence}

The XS10 GRBs sample is made out by collecting the data available in the literature so that the final catalog is not homogenous at all. In particular, the satellite used to get both the spectral and afterglows data change from one case to another. In order to investigate whether this could have any impact on the HD, we consider again the deviations from the fiducial $\Lambda$CDM model using only the 80 GRBs detected with the {\it Swift} satellite (excluding outliers). We get\,:

\begin{displaymath}
\langle \Delta \mu \rangle = -0.29 \ \ , \ \ (\Delta \mu)_{med} = 0.03 \ \ , \ \ (\Delta \mu)_{rms} = 1.33 \ \ ,
\end{displaymath}

\begin{displaymath}
\langle \Delta \mu \rangle = -0.24 \ \ , \ \ (\Delta \mu)_{med} = 0.11 \ \ , \ \ (\Delta \mu)_{rms} = 1.32 \ \ ,
\end{displaymath}

\begin{displaymath}
\langle \Delta \mu \rangle = -0.40 \ \ , \ \ (\Delta \mu)_{med} = 0.05 \ \ , \ \ (\Delta \mu)_{rms} = 1.40 \ \ ,
\end{displaymath}
including all correlations and using the $F$, $C$ and $LR$ samples, respectively, while we find

\begin{displaymath}
\langle \Delta \mu \rangle = -0.25 \ \ , \ \ (\Delta \mu)_{med} = -0.05 \ \ , \ \ (\Delta \mu)_{rms} = 1.18 \ \ ,
\end{displaymath}

\begin{displaymath}
\langle \Delta \mu \rangle = -0.18 \ \ , \ \ (\Delta \mu)_{med} = 0.06 \ \ , \ \ (\Delta \mu)_{rms} = 1.18 \ \ ,
\end{displaymath}

\begin{displaymath}
\langle \Delta \mu \rangle = -0.33 \ \ , \ \ (\Delta \mu)_{med} = -0.05 \ \ , \ \ (\Delta \mu)_{rms} = 1.24 \ \ ,
\end{displaymath}
excluding the $L$\,-\,$E_{peak}$ and $L$\,-\,$V$ correlations and the same samples. Somewhat surprisingly, we find larger $\Delta \mu$ values independent on the calibration procedure adopted and use of all or only four out of the six scaling laws. Actually, this is most likely a consequence of the different redshift range probed. Indeed, $\Delta \mu$ positively correlates with $z$ so that a sample with a larger median redshift will have a larger $\langle \Delta \mu \rangle$. This is indeed the case here since the pre\,-\,{\it Swift} data are limited to a smaller redshift range mainly because of instrumental effects. We therefore agree with the results in XS10 who have argued against a systematic difference among pre and post\,-\,{\it Swift} GRBs datasets.

\section{Conclusions}

GRBs have recently attracted a lot of attention as promising candidates to expand the Hubble diagram up to very high $z$, deep into the matter dominated era thus complementing SNeIa which are, on the contrary, excellent probes of the dark energy epoch. However, still much work is needed in order to be sure that GRBs can indeed hold this promise.

As the Phillips law is the basic tool to standardize SNeIa, the hunt for a similar relation to be used for GRBs has lead to different empirically motivated 2D scaling relations. However, the lack of a local GRBs sample leads to the so called {\it circularity problem}, i.e. the need to know the cosmological model to infer the luminosity distance to each GRB contrasted with the desire to constrain that same cosmological model. In an attempt to overcome this problem, we have here considered the impact on the scaling relations and GRBs HD of three different methods to estimate the luminosity distance. First, we rely on a fiducial $\Lambda$CDM model fitted to the SNeIa data. Then, we make a step further towards a model independent calibration using the cosmographic expansion to the fifth order in $z$ thus only assuming that the Robertson\,-\,Walker metric describe the universe background. Finally, we use local regression on SNeIa to interpolate the luminosity distance to a given $z$ with no assumption at all on the cosmological model. We find that these three conceptually different methods to estimate the luminosity distance and hence calibrate the GRBs scaling relations lead to consistent results. Indeed, the slope, zeropoint and intrinsic scatter of the 2D correlations are in good agreement within the $68\%$ confidence ranges. As a consequence, the Hubble diagrams averaging over the six correlations considered is not affected by the choice of the calibration method being the distance moduli to each GRB in agreement within the errors. Such a preliminary conclusion may suggest that the calibration problem could actually be a false problem since the use of a fiducial model to estimate the luminosity distance to the $z \le 1.4$ GRBs finally lead to the same HD of a fully model independent method such as the local regression. Actually, some important caveats have to be taken into account to not overrate this result. First, we have chosen as a fiducial model the concordance $\Lambda$CDM one which is known to fit very well the SNeIa data. As such, the predicted luminosity distance closely follows the one inferred from SNeIa which can be estimated using local regression. As such, these two apparently different luminosity distance methods actually give the same luminosity distance values so that the agreement of the constraints on the GRBs parameters is not surprising. Should we have used as fiducial model a different dark energy scenario, the results could have been different. Moreover, although not statistically meaningful, a weak dependence of the slope of the GRBs scaling relations on the calibration method is indeed present with the local regression leading to shallower relations. A larger GRBs sample is therefore needed to check whether these trends are only a statistically meaningless fluctuation or an evidence of the key role of the calibration method adopted.

Once the calibration procedure has been adopted, one has still to check whether a redshift evolution of the GRBs scaling relations is present or not. Should indeed the scaling relations coefficients be a function of $z$, neglecting such a dependence would bias both the calibration and the Hubble diagram thus introducing a systematic error when using the derived HD to constrain the cosmological parameters. We have therefore explored two different parameterizations assuming that the zeropoint only changes with $z$ or that both the slope and the zeropoint may be approximated as linear functions of the redshift. Although there are some evidences in favor of a significant evolution of the zeropoint with the redshift, it is nevertheless worth stressing that adding one or two more parameters to the fit of the scaling relations does not improve the quality of the fit (the reduced $\chi^2$ being the same or larger) and do not reduce the intrinsic scatter. The Occam's razor therefore makes us argue that such an evolution is not statistically motivated and neglecting it is actually a conservative and better motivated choice. It is, however, worth saying that such a conclusion heavily relies on the quality of the data. Indeed, the value of the $\chi^2$ depends also on the size of the uncertainties. For instance, artificially reducing by hand the errors leaving unchanged the central values increases the statistical significance of the redshift evolution making this choice the most conservative one. On the other hand, one should reconsider this issue by preliminarily investigating whether a selection effect is present. Indeed, we have implicitly assumed that the probability to measure the quantities entering the fitted correlation is the same whatever the GRB redshift is. Should this not be the case, one has to include a prior in the likelihood function to model this selection effect and check whether the results are affected or not by the prior itself. However, such an analysis can only be made on a case\,-\,by\,-\,case basis and asks for a preliminary theoretical modeling of the correlation of interest to simulate the full detection process.

Assuming that no evolution is present, we have finally checked that the derived Hubble diagrams are not affected by systematics related to the choice of the calibration method, the averaging procedure or the homogeneity of the sample. As such, the GRBs HD could be safely used as a tool to constrain cosmological parameters. Although these results are strongly encouraging, we however warn the reader that they are still preliminary. Indeed, what we have actually shown is that the systematics induced by the different effects we have considered are smaller than the statistical errors due to measurement uncertainties of the GRBs observable quantities and the intrinsic scatter of the 2D scaling relations used. It is not possible to forecast whether the conclusions still hold should these errors be reduced. To this end, increasing the number of GRBs and improving the precision could not be the right strategy. Indeed, the error budget on the distance modulus is typically dominated by the intrinsic scatter so that one should rather rely on GRBs scaling relations which are as tight as possible. From this point of view, for instance, one should give off the $L$\,-\,$E_{peak}$ and $L$\,-\,$V$ correlations, but this comes at the price of reducing the number of usable GRBs and increase the error (since we now average on a lower number of $\mu$ estimates). Again, this leads to the need of larger GRBs samples to select the tightest correlations without degrading the precision on the distance modulus determination.

As a final remark, an analogy between SNeIa and GRBs as cosmological tools can be considered. As soon as the Phillips law was established, the SNeIa started to be used to probe the Hubble diagram up to $z \sim 1.5$ notwithstanding possible problems with the evolution of the standardization method adopted and the universality of its basic assumptions. Nowadays, the SNeIa samples have so many objects that it is possible not only to use them as cosmological tools, but also carefully explore systematics and their impact on the Hubble diagram. The present day situation for GRBs is similar with many Phillips law\,-\,like relations proposed to standardize them and most attention dedicated more to their cosmological use than to the systematics. As for SNeIa, we must therefore only wait that, as time goes by and new instruments bring us larger and higher quality samples, the issue of GRBs systematics can be more efficiently faced off showing that these powerful explosions can indeed hold the promise to fill the gap between the dark energy era probed by SNeIa and the early universe tested by CMBR. \\

{\it Acknowledgements} VFC is supported by the Italian Space Agency (ASI).

\appendix

\section*{Appendix A}

We report here for completeness the constraints on the calibration parameters for the two different redshift evolution parameterizations and the three luminosity distance estimate methods. Although the best fit values may be different from one case to another, the confidence ranges for the parameters always overlap quite well so that the difference is not statistically meaningful. The main conclusions in Sect.\,4 qualitatively apply also changing the calibration method so that we do not repeat here the analysis of the results.

\begin{table*}
\caption{Constraints on the calibration parameters $(a, b, \alpha, \sigma_{int})$ for the 2D correlations considered in the text and fitted to the $F$, $C$ and $LR$ samples (first, second and third row of each fit). Columns are as follows\,: 1. correlation id, 2. reduced $\chi^2$, 3. best fit parameters, 4., 5., 6., 7. median values and $68\%$ confidence ranges for the fit parameters. The number of GRBs used in the same as in Table\,\ref{tab: calpar}.}
\label{tab: plparall}
\begin{center}
\begin{tabular}{|c|c|c|c|c|c|c|}
\hline
Id & $\chi^2/d.o.f.$ & $(a, b, \alpha, \sigma_{int})_{bf}$ & $a$ & $b$ & $\alpha$ & $\sigma_{int}$ \\
\hline \hline

~ & ~ & ~ & ~ & ~ & ~ & ~    \\

$L$\,-\,$E_{peak}$ & 1.08 & (0.89, 50.77, 0.67, 0.41) & $0.82_{-0.19}^{+0.20}$ & $50.75_{-0.65}^{+0.59}$ & $0.68_{-0.30}^{+0.18}$ & $0.44_{-0.06}^{+0.06}$ \\

~ & ~ & ~ & ~ & ~ & ~ & ~    \\

$L$\,-\,$E_{peak}$ & 1.13 & (0.90, 50.84, 0.61, 0.40) & $0.83_{-0.19}^{+0.22}$ & $50.84_{-0.60}^{+0.55}$ & $0.62_{-0.29}^{+0.32}$ & $0.43_{-0.05}^{+0.07}$ \\

~ & ~ & ~ & ~ & ~ & ~ & ~    \\

$L$\,-\,$E_{peak}$ & 1.17 & (0.75, 50.88, 0.65, 0.38) & $0.66_{-0.17}^{+0.19}$ & $50.78_{-0.65}^{+0.63}$ & $0.72_{-0.34}^{+0.34}$ & $0.41_{-0.06}^{+0.05}$ \\

~ & ~ & ~ & ~ & ~ & ~ & ~    \\
\hline
~ & ~ & ~ & ~ & ~ & ~ & ~    \\

$L$\,-\,$\tau_{lag}$ & 1.10 & (-0.58, 50.38, 0.88, 0.39) & $-0.57_{-0.14}^{+0.12}$ & $50.39_{-0.58}^{+0.66}$ & $0.87_{-0.33}^{+0.30}$ & $0.41_{-0.06}^{+0.07}$ \\

~ & ~ & ~ & ~ & ~ & ~ & ~    \\

$L$\,-\,$\tau_{lag}$ & 1.19 & (-0.58, 50.39, 0.85, 0.38) & $-0.54_{-0.13}^{+0.13}$ & $50.31_{-0.56}^{+0.57}$ & $0.89_{-0.28}^{+0.29}$ & $0.41_{-0.06}^{+0.07}$ \\

~ & ~ & ~ & ~ & ~ & ~ & ~    \\

$L$\,-\,$\tau_{lag}$ & 1.24 & (-0.55, 50.45, 0.88, 0.31) & $-0.52_{-0.12}^{+0.11}$ & $50.38_{-0.50}^{+0.53}$ & $0.91_{-0.27}^{+0.25}$ & $0.33_{-0.05}^{+0.07}$ \\

~ & ~ & ~ & ~ & ~ & ~ & ~    \\
\hline
~ & ~ & ~ & ~ & ~ & ~ & ~    \\

$L$\,-\,$\tau_{RT}$ & 1.16 & (-0.79, 51.22, 0.65, 0.44) & $-0.64_{-0.25}^{+0.21}$ & $50.97_{-0.75}^{+0.70}$ & $0.76_{-0.33}^{+0.36}$ & $0.47_{-0.06}^{+0.07}$ \\

~ & ~ & ~ & ~ & ~ & ~ & ~    \\

$L$\,-\,$\tau_{RT}$ & 1.13 & (-0.72, 51.08, 0.69, 0.45) & $-0.63_{-0.22}^{+0.18}$ & $50.97_{-0.66}^{+0.64}$ & $0.73_{-0.30}^{+0.34}$ & $0.47_{-0.06}^{+0.08}$ \\

~ & ~ & ~ & ~ & ~ & ~ & ~    \\

$L$\,-\,$\tau_{RT}$ & 1.20 & (-0.76, 51.32, 0.64, 0.36) & $-0.66_{-0.24}^{+0.20}$ & $51.20_{-0.80}^{+0.71}$ & $0.68_{-0.34}^{+0.38}$ & $0.40_{-0.06}^{+0.07}$ \\

~ & ~ & ~ & ~ & ~ & ~ & ~    \\

\hline

~ & ~ & ~ & ~ & ~ & ~ & ~    \\

$L$\,-\,$V$ & 1.23 & (-0.10, 50.27, 1.07, 0.54) & $-0.05_{-0.37}^{+0.26}$ & $50.29_{-0.96}^{+0.91}$ & $1.04_{-0.50}^{+0.52}$ & $0.57_{-0.08}^{+0.11}$ \\

~ & ~ & ~ & ~ & ~ & ~ & ~    \\

$L$\,-\,$V$ & 1.23 & (-0.14, 50.31, 1.05, 0.54) & $-0.09_{-0.37}^{+0.16}$ & $50.26_{-0.82}^{+0.82}$ & $1.07_{-0.46}^{+0.47}$ & $0.57_{-0.08}^{+0.10}$ \\

~ & ~ & ~ & ~ & ~ & ~ & ~    \\

$L$\,-\,$V$ & 1.36 & (-0.10, 50.32, 1.10, 0.42) & $-0.10_{-0.30}^{+0.12}$ & $50.38_{-0.76}^{+0.70}$ & $1.07_{-0.42}^{+0.41}$ & $0.46_{-0.07}^{+0.09}$ \\

~ & ~ & ~ & ~ & ~ & ~ & ~    \\
\hline

~ & ~ & ~ & ~ & ~ & ~ & ~    \\

$E_{\gamma}$\,-\,$E_{peak}$ & 1.37 & (1.57, 50.64, 0.00, 0.14) & $1.48_{-0.26}^{+0.23}$ & $50.42_{-0.57}^{+0.51}$ & $0.13_{-0.28}^{+0.31}$ & $0.20_{-0.06}^{+0.07}$ \\

~ & ~ & ~ & ~ & ~ & ~ & ~    \\

$E_{\gamma}$\,-\,$E_{peak}$ & 1.38 & (1.60, 50.66, -0.01, 0.14) & $1.44_{-0.29}^{+0.24}$ & $50.34_{-0.68}^{+0.48}$ & $0.16_{-0.25}^{+0.37}$ & $0.21_{-0.07}^{+0.08}$ \\

~ & ~ & ~ & ~ & ~ & ~ & ~    \\

$E_{\gamma}$\,-\,$E_{peak}$ & 1.35 & (1.58, 50.74, -0.03, 0.12) & $1.47_{-0.25}^{+0.24}$ & $50.46_{-0.61}^{+0.52}$ & $0.12_{-0.29}^{+0.33}$ & $0.19_{-0.08}^{+0.09}$ \\

~ & ~ & ~ & ~ & ~ & ~ & ~    \\
\hline
~ & ~ & ~ & ~ & ~ & ~ & ~    \\

$E_{iso}$\,-\,$E_{peak}$ & 0.81 & (1.39, 51.58, 0.58, 0.48) & $1.29_{-0.28}^{+0.28}$ & $51.57_{-0.81}^{+0.84}$ & $0.58_{-0.45}^{+0.44}$ & $0.50_{-0.08}^{+0.10}$ \\

~ & ~ & ~ & ~ & ~ & ~ & ~    \\

$E_{iso}$\,-\,$E_{peak}$ & 0.81 & (1.40, 51.70, 0.50, 0.48) & $1.28_{-0.29}^{+0.28}$ & $51.43_{-0.81}^{+0.75}$ & $0.64_{-0.40}^{+0.43}$ & $0.50_{-0.08}^{+0.10}$ \\

~ & ~ & ~ & ~ & ~ & ~ & ~    \\

$E_{iso}$\,-\,$E_{peak}$ & 0.80 & (1.41, 51.98, 0.37, 0.47) & $1.32_{-0.30}^{+0.28}$ & $51.89_{-0.86}^{+0.75}$ & $0.42_{-0.39}^{+0.46}$ & $0.50_{-0.08}^{+0.10}$ \\

~ & ~ & ~ & ~ & ~ & ~ & ~    \\
\hline
\end{tabular}
\end{center}
\end{table*}

\begin{table*}
\caption{Constraints on the calibration parameters $(a_0, \log{a_1}, b_0, \log{b_1}, \sigma_{int})$ for the 2D correlations considered in the text and fitted to the $F$, $C$ and $LR$ samples (first, second and third row of each fit). Columns are as follows\,: 1. correlation id, 2. reduced $\chi^2$, 3. best fit parameters, 4., 5., 6., 7., 8. median values and $68\%$ confidence ranges for the fit parameters. Number of GRBs as in Table\,\ref{tab: calpar}.}
\label{tab: linparall}
\begin{center}
\begin{tabular}{|c|c|c|c|c|c|c|c|}
\hline
Id & $\chi^2/d.o.f.$ & $(a_0, \log{a_1}, b_0, \log{b_1}, \sigma_{int})_{bf}$ & $a_0$ & $\log{a_1}$ & $b_0$ & $\log{b_1}$ & $\sigma_{int}$ \\
\hline \hline

~ & ~ & ~ & ~ & ~ & ~ & ~ & ~    \\

$L$\,-\,$E_{peak}$ & 1.12 & (0.89, -2.69, 51.49, -0.21, 0.41) & $0.89_{-0.25}^{+0.22}$ & $-3.05_{-3.63}^{+2.13}$ & $51.87_{-0.46}^{+0.19}$ & $-0.78_{-3.06}^{+0.64}$ & $0.44_{-0.05}^{+0.03}$ \\

~ & ~ & ~ & ~ & ~ & ~ & ~ & ~    \\

$L$\,-\,$E_{peak}$ & 1.13 & (0.62, -0.57, 51.44, -0.20, 0.41) & $0.84_{-0.27}^{+0.23}$ & $-2.83_{-4.19}^{+2.67}$ & $51.83_{-0.44}^{+0.20}$ & $-0.71_{-3.02}^{+0.55}$ & $0.44_{-0.06}^{+0.06}$ \\

~ & ~ & ~ & ~ & ~ & ~ & ~ & ~    \\

$L$\,-\,$E_{peak}$ & 1.18 & (0.73, -2.71, 51.50, -0.16, 0.38) & $0.75_{-0.23}^{+0.21}$ & $-2.76_{-3.39}^{+1.75}$ & $52.00_{-0.50}^{+0.16}$ & $-1.05_{-3.76}^{+0.89}$ & $0.41_{-0.05}^{+0.06}$ \\

~ & ~ & ~ & ~ & ~ & ~ & ~ & ~    \\

\hline

~ & ~ & ~ & ~ & ~ & ~ & ~ & ~    \\

$L$\,-\,$\tau_{lag}$ & 1.13 & (-0.60, -4.43, 51.31, -0.07, 0.40) & $-0.64_{-0.19}^{+0.16}$ & $-2.78_{-2.85}^{+1.61}$ & $51.61_{-0.45}^{+0.49}$ & $-0.28_{-2.83}^{+0.27}$ & $0.43_{-0.06}^{+0.08}$ \\

~ & ~ & ~ & ~ & ~ & ~ & ~ & ~    \\

$L$\,-\,$\tau_{lag}$ & 1.21 & (-0.61, -2.61, 51.25, -0.07, 0.39) & $-0.59_{-0.14}^{+0.13}$ & $-3.13_{-3.50}^{+1.72}$ & $51.43_{-0.37}^{+0.50}$ & $-0.17_{-0.82}^{+0.18}$ & $0.42_{-0.06}^{+0.08}$ \\

~ & ~ & ~ & ~ & ~ & ~ & ~ & ~    \\

$L$\,-\,$\tau_{lag}$ & 1.30 & (-0.74, -0.60, 51.27, -0.02, 0.31) & $-0.58_{-0.14}^{+0.13}$ & $-2.98_{-4.55}^{+1.84}$ & $51.47_{-0.32}^{+0.45}$ & $-0.13_{-0.42}^{+0.11}$ & $0.34_{-0.06}^{+0.08}$ \\

~ & ~ & ~ & ~ & ~ & ~ & ~ & ~    \\

\hline

~ & ~ & ~ & ~ & ~ & ~ & ~ & ~    \\

$L$\,-\,$\tau_{RT}$ & 1.17 & (-1.56, -0.08, 52.38, -0.70, 0.44) & $-0.95_{-0.46}^{+0.33}$ & $-1.81_{-3.90}^{+1.59}$ & $52.42_{-0.53}^{+0.19}$ & $-1.81_{-3.91}^{+1.60}$ & $0.47_{-0.06}^{+0.07}$ \\

~ & ~ & ~ & ~ & ~ & ~ & ~ & ~    \\

$L$\,-\,$\tau_{RT}$ & 1.22 & (-1.59, -0.07, 52.29, -0.63, 0.44) & $-0.82_{-0.35}^{+0.27}$ & $-2.36_{-3.33}^{+1.86}$ & $52.28_{-0.60}^{+0.23}$ & $-0.92_{-3.36}^{+0.79}$ & $0.47_{-0.06}^{+0.08}$ \\

~ & ~ & ~ & ~ & ~ & ~ & ~ & ~    \\

$L$\,-\,$\tau_{RT}$ & 1.27 & (-1.83, 0.07, 52.60, -2.77, 0.33) & $-1.09_{-0.60}^{+0.17}$ & $-0.89_{-3.90}^{+0.89}$ & $52.54_{-0.35}^{+0.16}$ & $-2.43_{-4.07}^{+2.04}$ & $0.39_{-0.06}^{+0.07}$ \\

~ & ~ & ~ & ~ & ~ & ~ & ~ & ~    \\
\hline

~ & ~ & ~ & ~ & ~ & ~ & ~ & ~    \\

$L$\,-\,$V$ & 1.36 & (-0.15, -2.16, 51.40, 0.03, 0.52) & $-0.10_{-0.44}^{+0.41}$ & $-1.79_{-4.81}^{+1.49}$ & $51.85_{-0.53}^{+0.39}$ & $-0.77_{-2.86}^{+0.76}$ & $0.58_{-0.08}^{+0.10}$ \\

~ & ~ & ~ & ~ & ~ & ~ & ~ & ~    \\

$L$\,-\,$V$ & 1.24 & (-0.11, -2.68, 51.32, 0.02, 0.56) & $-0.21_{-0.47}^{+0.44}$ & $-2.15_{-3.78}^{+1.49}$ & $51.85_{-0.53}^{+0.39}$ & $-0.77_{-2.86}^{+0.76}$ & $0.58_{-0.08}^{+0.10}$ \\

~ & ~ & ~ & ~ & ~ & ~ & ~ & ~    \\

$L$\,-\,$V$ & 1.40 & (-0.13, -3.13, 51.40, 0.06, 0.43) & $-0.02_{-0.44}^{+0.40}$ & $-1.94_{-2.39}^{+1.56}$ & $51.86_{-0.46}^{+0.41}$ & $-0.57_{-3.83}^{+0.59}$ & $0.49_{-0.08}^{+0.10}$ \\

~ & ~ & ~ & ~ & ~ & ~ & ~ & ~    \\
\hline

~ & ~ & ~ & ~ & ~ & ~ & ~ & ~    \\

$E_{\gamma}$\,-\,$E_{peak}$ & 1.70 & (0.79, 0.04, 50.37, -0.56, 0.10) & $1.40_{-0.47}^{+0.23}$ & $-1.70_{-2.67}^{+1.63}$ & $50.61_{-0.15}^{+0.08}$ & $-2.86_{-3.84}^{+1.91}$ & $0.17_{-0.06}^{+0.08}$ \\

~ & ~ & ~ & ~ & ~ & ~ & ~ & ~    \\

$E_{\gamma}$\,-\,$E_{peak}$ & 1.49 & (0.98, -0.05, 50.46, -0.92, 0.12) & $1.43_{-0.49}^{+0.22}$ & $-2.42_{-4.22}^{+2.36}$ & $50.59_{-0.12}^{+0.08}$ & $-2.98_{-3.35}^{+2.91}$ & $0.17_{-0.06}^{+0.08}$ \\

~ & ~ & ~ & ~ & ~ & ~ & ~ & ~    \\

$E_{\gamma}$\,-\,$E_{peak}$ & 1.70 & (0.74, 0.09, 50.40, -0.57, 0.06) & $1.47_{-0.37}^{+0.19}$ & $-2.70_{-4.51}^{+2.41}$ & $50.64_{-0.12}^{+0.08}$ & $-2.90_{-4.00}^{+1.69}$ & $0.16_{-0.07}^{+0.09}$ \\

~ & ~ & ~ & ~ & ~ & ~ & ~ & ~    \\

\hline

~ & ~ & ~ & ~ & ~ & ~ & ~ & ~    \\

$E_{iso}$\,-\,$E_{peak}$ & 0.79 & (0.94, -0.23, 52.00, -0.15, 0.49) & $1.27_{-0.50}^{+0.33}$ & $-1.78_{-2.54}^{+1.61}$ & $52.56_{-0.34}^{+0.16}$ & $-1.73_{-2.41}^{+1.42}$ & $0.50_{-0.08}^{+0.10}$ \\

~ & ~ & ~ & ~ & ~ & ~ & ~ & ~    \\

$E_{iso}$\,-\,$E_{peak}$ & 0.91 & (0.69, -0.07, 51.98, -0.12, 0.44) & $1.20_{-0.92}^{+0.39}$ & $-1.36_{-3.41}^{+1.47}$ & $52.53_{-0.43}^{+0.16}$ & $-1.93_{-3.31}^{+1.67}$ & $0.49_{-0.07}^{+0.08}$ \\

~ & ~ & ~ & ~ & ~ & ~ & ~ & ~    \\

$E_{iso}$\,-\,$E_{peak}$ & 0.86 & (0.66, -0.05, 52.21, -0.28, 0.45) & $1.24_{-0.72}^{+0.34}$ & $-1.85_{-2.98}^{+1.85}$ & $52.57_{-0.39}^{+0.17}$ & $-1.84_{-2.80}^{+1.56}$ & $0.49_{-0.08}^{+0.10}$ \\

~ & ~ & ~ & ~ & ~ & ~ & ~ & ~    \\
\hline
\end{tabular}
\end{center}
\end{table*}

\end{document}